\documentstyle[epsfig,12pt]{article}
\input{psfig.sty}
\newlength{\dinwidth}
\newlength{\dinmargin}
\newlength{\extraspace}
\newlength{\extraspaces}
\newcommand{\be}{\begin{equation}
\addtolength{\abovedisplayskip}{\extraspaces}
\addtolength{\belowdisplayskip}{\extraspaces}
\addtolength{\abovedisplayshortskip}{\extraspace}
\addtolength{\belowdisplayshortskip}{\extraspace}}
\newcommand{\ee}{\end{equation}}
\newcommand{\bdm}{\begin{displaymath}
\addtolength{\abovedisplayskip}{\extraspaces}
\addtolength{\belowdisplayskip}{\extraspaces}
\addtolength{\abovedisplayshortskip}{\extraspace}
\addtolength{\belowdisplayshortskip}{\extraspace}}
\newcommand{\edm}{\end{displaymath}}
\def\simlt{\mathrel{\lower2.5pt\vbox{\lineskip=0pt\baselineskip=0pt
           \hbox{$<$}\hbox{$\sim$}}}}
\def\simgt{\mathrel{\lower2.5pt\vbox{\lineskip=0pt\baselineskip=0pt
           \hbox{$>$}\hbox{$\sim$}}}}
\newcommand{\ls}[1]
   {\dimen0=\fontdimen6\the\font
    \lineskip=#1\dimen0
    \advance\lineskip.5\fontdimen5\the\font
    \advance\lineskip-\dimen0
    \lineskiplimit=.9\lineskip
    \baselineskip=\lineskip
    \advance\baselineskip\dimen0
    \normallineskip\lineskip
    \normallineskiplimit\lineskiplimit
    \normalbaselineskip\baselineskip
    \ignorespaces}

\catcode`@=11
\newcount\@tempcntc
\def\@citex[#1]#2{\if@filesw\immediate\write\@auxout{\string\citation{#2}}\fi
  \@tempcnta\z@\@tempcntb\m@ne\def\@citea{}\@cite{\@for\@citeb:=#2\do
    {\@ifundefined
       {b@\@citeb}{\@citeo\@tempcntb\m@ne\@citea\def\@citea{,}{\bf ?}\@warning
       {Citation `\@citeb' on page \thepage \space undefined}}%
    {\setbox\z@\hbox{\global\@tempcntc0\csname b@\@citeb\endcsname\relax}%
     \ifnum\@tempcntc=\z@ \@citeo\@tempcntb\m@ne
       \@citea\def\@citea{,}\hbox{\csname b@\@citeb\endcsname}%
     \else
      \advance\@tempcntb\@ne
      \ifnum\@tempcntb=\@tempcntc
      \else\advance\@tempcntb\m@ne\@citeo
      \@tempcnta\@tempcntc\@tempcntb\@tempcntc\fi\fi}}\@citeo}{#1}}
\def\@citeo{\ifnum\@tempcnta>\@tempcntb\else\@citea\def\@citea{,}%
  \ifnum\@tempcnta=\@tempcntb\the\@tempcnta\else
   {\advance\@tempcnta\@ne\ifnum\@tempcnta=\@tempcntb \else \def\@citea{--}\fi
    \advance\@tempcnta\m@ne\the\@tempcnta\@citea\the\@tempcntb}\fi\fi}
\catcode`@=12

\newcommand{\Tr}{\hbox{\rm Tr}}
\def\simge{\mathrel{%
   \rlap{\raise 0.511ex \hbox{$>$}}{\lower 0.511ex \hbox{$\sim$}}}}
\def\simle{\mathrel{
   \rlap{\raise 0.511ex \hbox{$<$}}{\lower 0.511ex \hbox{$\sim$}}}}
 
\def\slashchar#1{\setbox0=\hbox{$#1$}           
   \dimen0=\wd0                                 
   \setbox1=\hbox{/} \dimen1=\wd1               
   \ifdim\dimen0>\dimen1                        
      \rlap{\hbox to \dimen0{\hfil/\hfil}}      
      #1                                        
   \else                                        
      \rlap{\hbox to \dimen1{\hfil$#1$\hfil}}   
      /                                         
   \fi}                                         %
\def\nn{\nonumber}
\def\ts{\thinspace}

\def\ra{\rightarrow}

\def\lra{\longrightarrow}

\def\ol{\bar}
\def\be{\begin{equation}} 
\def\ee{\end{equation}} 
\def\bea{\begin{eqnarray}}
\def\eea{\end{eqnarray}}
\def\ba{\begin{array}}
\def\ea{\end{array}}
\def\dag{\dagger}

\def\CH{{\cal H}}

\def\CL{{\cal L}}
\def\CM{{\cal M}}

\def\CO{{\cal O}}

\def\CW{{\cal W}}

\def\kslash{\raise.15ex\hbox{/}\kern-.57em k}

\def\half{{\textstyle{ { 1\over { 2 } }}}}
\def\threehalves{{\textstyle{ { 3\over { 2 } }}}}

\def\nin{\noindent}

\begin{document}
\title{
\vskip -15mm
\begin{flushright}
\vskip -15mm
{\small FERMILAB-PUB-02/019-T\\
BUHEP-02-11\\
hep-ph/0202093\\}
\vskip 5mm
\end{flushright}
{\Large{\bf \hskip 0.38truein
A Case Study in Dimensional Deconstruction}}\\
}
\author{
\centerline{{Kenneth Lane\thanks{lane@physics.bu.edu}}}\\
\centerline{{Fermi National Accelerator Laboratory}}\\
\centerline{{P.O. Box 500, Batavia, IL 60510}}\\
\centerline{and}\\
\centerline{{Department of Physics, Boston University}}\\
\centerline{{590 Commonwealth Avenue, Boston, MA 02215\footnote{Permanent
address.}}}\\
}
\maketitle
\begin{abstract}

We test Arkani-Hamed et al.'s dimensional deconstruction on a model that is
predicted to have a naturally light composite Higgs boson, i.e., one whose
mass $M$ is much less than its binding scale $\Lambda$, and whose quartic
coupling $\lambda$ is large, so that its vacuum expectation value $v \sim
M/\sqrt{\lambda} \ll \Lambda$ also. We consider two different underlying
dynamics---UV completions---at the scale $\Lambda$ for this model. We find
that the expectation from dimensional deconstruction is not realized and that
low energy details depend crucially on the UV completion. In one case, $M \ll
\Lambda$ and $\lambda \ll 1$, hence, $v \sim \Lambda$. In the other,
$\lambda$ can be large or small, but then so is $M$, and $v$ is still
$\CO(\Lambda)$.

\end{abstract}


\newpage

\begin{figure}[tb]
\vbox to 9cm{
\vfill
\includegraphics{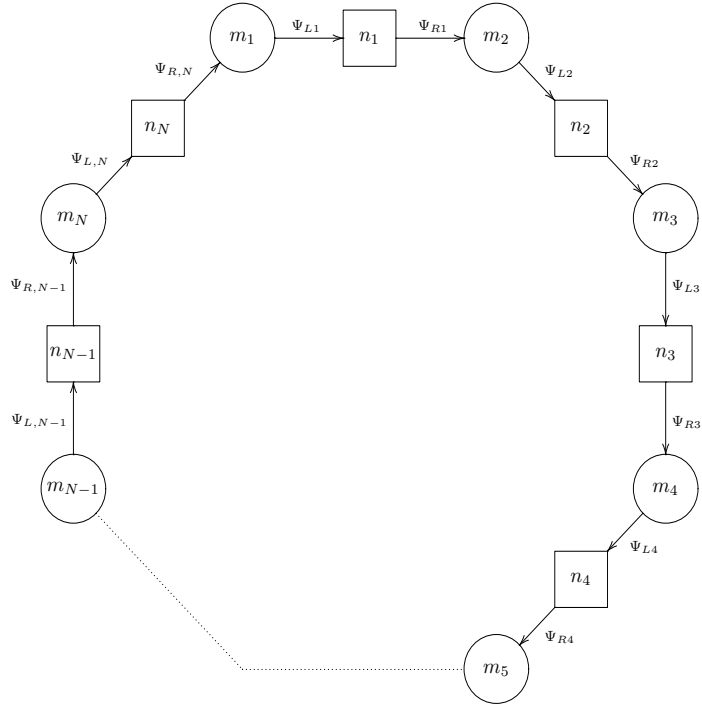}
\vfill
}
\caption{The full moose for the ring model of Ref.~\cite{acga}, showing its
  UV completion. Strong gauge goups are labeled by $n_1,n_2,\dots,n_N$ and
  weak gauge groups by $m_1,m_2,\dots,m_N$. Fermions $\psi_{Lk}$ and
  $\psi_{Rk}$ transform as $(n,m,1)$ and $(n,1,m)$ of $(SU(n)_k
  \otimes SU(m)_k \otimes SU(m)_{k+1})$.
\label{fig:a}}
\end{figure}


\begin{figure}[tb]
\vbox to 6.5cm{
\vfill
\includegraphics{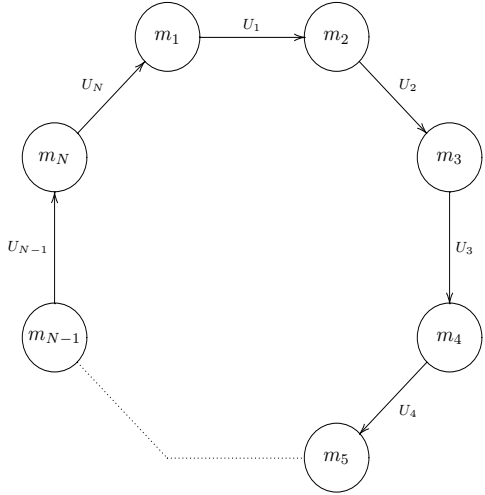}
\vfill
}
\caption{The condensed moose for the ring model of Ref.~\cite{acga},
  characterizing its low--energy structure with nonlinear sigma fields $U_k =
  \exp{(i\pi_k/f)}$ linking the weak groups $SU(m)_k$ and $SU(m)_{k+1}$.
\label{fig:a}}
\end{figure}


\begin{figure}[tb]
\vbox to 9cm{
\vfill
\includegraphics{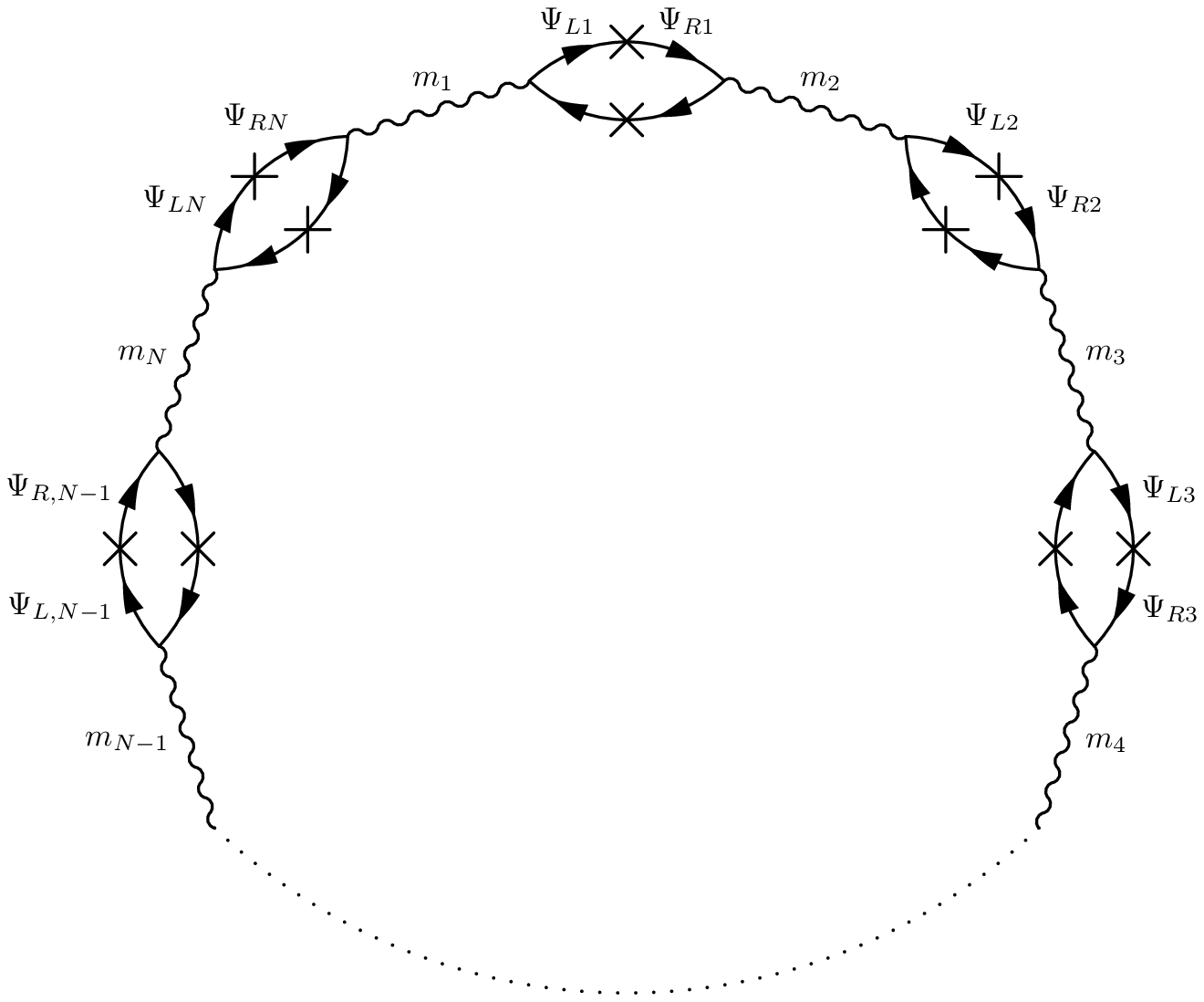}
\vfill
}
\caption{Graphical depiction of the Hamiltonian $\CH_N$ in
  Eq.~(\ref{eq:HN}). Fermions transform as indicated in
  Eq.~(\ref{eq:fermionsa}) under the weak gauge groups $SU(m)_k$, whose bosons
  are identified in the figure. An $\times$~indicates a dynamical mass
  insertion. Strong $SU(n)$ gauge boson interactions within each fermion loop
  are not indicated. There are no strong gauge interactions between loops.
\label{fig:a}}
\end{figure}


\section*{1. Introduction}

There has been considerable interest lately in a new approach to
model--building called ``dimensional deconstruction''. There are two views of
dimensional deconstruction. One, taken by Arkani-Hamed, Cohen and Georgi
(ACG)~\cite{acga,acgb} is that certain 4--dimensional theories look, for a
range of energies, like higher dimensional theories in which the compactified
extra dimensions are discretized on a periodic lattice. ACG used this
resemblance---particularly the topological similarity between the $d > 4$
components of gauge fields and certain 4--dimensional Goldstone bosons, and
the absence of divergent counterterms for gauge--invariant operators of
dimension greater than $d$---to deduce the form, strength, and sensitivity to
high--scale physics of phenomenologically important operators such as mass
terms and self--interactions. The other view is that of Hill and his
collaborators~\cite{hilla,hillb,hillc} who assume the extra dimensions are
real. They discretize the extra dimensions too---to regulate the theory. This
``transverse lattice'' theory is expected to be in the same universality
class as the continuum theory. In the view of Hill et al. the connection
between gauge field components and light Higgs scalars is also
there---because they are the same thing---and so the allowed operators and
their sensitivity to high scale physics are unambiguous. The consequences of
both these views of dimensional deconstruction are similar, but they are not
identical.

In this paper we study ACG's view as they apply it to building a model of
electroweak symmetry breaking with light composite Higgs
bosons~\cite{lch}. In Ref.~\cite{acgb}, ACG used dimensional deconstruction
to deduce that certain pseudoGoldstone bosons (PGBs) acquire masses $M$ much
less than the energy scale at which they are formed, $\Lambda \simeq 4\pi f$,
where $f$ is the PGB decay constant. They argued further that the PGBs have
negative mass--squared terms $M_{-}^2 \sim -M^2$, and that their quartic
interaction is strong yet does not contribute to $M^2$. These
ingredients---positive and negative squared masses $M^2_{+} \simeq - M^2_{-}
\ll \Lambda^2$ and quartic couplings large compared to
$M^2_{\pm}/\Lambda^2$---are what's required for a light composite Higgs whose
vacuum expectation value $v \sim M$ {\it without} fine tuning. These PGBs are
prototypes for electroweak Higgs bosons whose mass and vev are {\it
naturally} stabilized far below their binding--energy scale. This is
important because it is the first natural scheme for electroweak symmetry
breaking since the inventions of technicolor and supersymmetry over 20 years
ago.

The simplest implementation of ACG's dimensional deconstruction for light
composite Higgses would be the naive one, which we dub the ``principal of
strict deconstruction'': For 4--dimensional theories which admit a higher
dimensional interpretation, the form and strength of operators involving
Goldstone bosons may be deduced from those for the corresponding $d > 4$
components of gauge fields. ACG certainly do not adopt such a strict
formulation for, as we will quickly see, it is incorrect. A more liberal
formulation is needed. In this paper we explore how much we must liberalize
it in order to achieve the goal of a naturally light composite Higgs. 

To that end, this paper is frankly pedagogical, containing many details of
the calculation of PGB masses and couplings. We hope that some will find the
pedagogy useful. For them, and for the experts, our bottom line is this:
Sometimes dimensional deconstruction works and sometimes it doesn't. It often
depends critically on the ultraviolet (UV) completion of the low--energy
theory to which deconstruction is applied.

To make this more concrete, let us review the simplest example presented by
ACG. In Ref.~\cite{acga} they introduced a model containing $N$ strong gauge
groups $SU(n)_k$ and $N$ weak ones $SU(m)_k$. The matter fields of this model
are the massless chiral fermions
\be\label{eq:fermionsa}
\psi_{Lk} \in (n,m,1)\ts, \ts\ts\ts \psi_{Rk} \in (n,1,m) \ts\ts\ts
{\rm of} \ts\ts\ts (SU(n)_k, \ts SU(m)_k, \ts SU(m)_{k+1}) \quad
(k=1,2,\dots,N) \ts.
\ee
The index $k$ is periodically identified with $k+N$, making this a ``moose
ring'' model depicted in Fig.~1. For simplicity, all $SU(n)$ couplings $g_s$
are taken equal. They become strong at the high energy scale $\Lambda$. All
$SU(m)$ couplings $g$ are taken equal and assumed to be much less than $g_s$
at $\Lambda$. This setup is the model's UV completion. Let us see how it
evolves as we descend to lower energies.

At $\Lambda$, the strong $SU(n)$ interactions cause the fermions to
condense as
\be\label{eq:svac}
\langle\Omega | \ol \psi_{Lk} \psi_{Rl} |\Omega\rangle = 
\langle\Omega | \ol \psi_{Rk} \psi_{Ll} |\Omega\rangle = -\delta_{kl}
\Delta\ts,
\ee
where $SU(m)$ indices are not summed over and $\Delta \simeq 4\pi f^3$. In
the limit $g\ra 0$, these fermions' interactions have a large chiral
symmetry, $[SU(m)_L \otimes SU(m)_R]^N$. The symmetry of the ground state
$|\Omega\rangle$ is the diagonal vectorial subgroup, $[SU(m)_V]^N$.
Therefore, there are $N$ sets of $m^2-1$ Goldstone bosons. They are the
pseudoscalars $\pi^a_k$ that couple to the axial currents $j^a_{5\mu,k}= \ol
\psi_{Rk} \gamma_\mu t_a \psi_{Rk} - \ol \psi_{Lk} \gamma_\mu t_a \psi_{Lk}$
with strength $2f$. Here, $t_a$ ($a = 1,2,\dots,m^2-1$) are generators in the
fundamental representation of $SU(m)$ normalized to $\Tr(t_a t_b) = \half
\delta_{ab}$.

Below $\Lambda$, this theory is described by nonlinear sigma model fields
$U_k = \exp{(i\pi_k^a t_a/f)} \equiv \exp{(i\pi_k/f)}$ interacting with
the weakly--coupled $SU(m)_k$ gauge fields $A_{k\mu} = A_{k\mu}^a t_a$. The
matter fields transform under the weak gauge groups as $U_k \ra W_k U_k
W^\dag_{k+1}$. The effective Lagrangian is
\be\label{eq:effLa}
\CL = - {1\over{2g^2}}\sum_{k=1}^N {\Tr} F_{k\ts\mu\nu} F_k^{\mu\nu} +
 f^2 \sum_{k=1}^N {\Tr} \left[(D_\mu U_k)^\dag D^\mu  U_k\right]\ts,
\ee
where $D_\mu U_k = \partial_\mu U_k -i A_{k\mu} U_k + i U_k
A_{k+1,\mu}$. This low energy theory is described by the ``condensed moose''
in Fig.~2, with the link variables $U_k$ connecting sites $k$ and $k+1$. In
this case, though not in all others, the mooses describing the high--energy
and low--energy theories look the same.

Now, $N-1$ of the gauge boson multiplets eat $N-1$ sets of Goldstone bosons
and acquire the masses $\CM_k = 2gf \sin(k \pi/N)$ for $k=1,\dots,N$. The
massless gauge field $A^a_\mu = (A^a_{1\mu} + \cdots + A^a_{N\mu})/\sqrt{N}$
couples with strength $g/\sqrt{N}$ and the uneaten Goldstone boson is $\pi^a=
(\pi^a_1 + \cdots + \pi^a_N)/\sqrt{N}$. In the unitary gauge, then, the
4--dimensional theory below $\Lambda$ is described by uniform link variables
$U_k =\exp{(i\pi^a t_a/\sqrt{N}f)}$ plus the massless and massive gauge
fields.

Alternatively, at energies well below $gf$, this looks exactly like a
5--dimensional gauge theory. The fifth dimension is compactified on a
discretized circle, represented exactly by the {\it condensed} moose, and
there appears to be (for $k\ll N$) a Kaluza--Klein tower of excitations of
the massless gauge boson~\cite{acga}. The circumference of the circle is
$R=Na$ where the lattice spacing $a = 1/gf$ and the 5--dimensional gauge
coupling is $g_5^2 = g^2 a$.  The fifth component of the gauge boson $A_5^a =
g\pi^a/\sqrt{N}$. The geometrical connection is clear: $\pi^a$ is the zero
mode associated with rotation about the circle of $SU(m)$ groups in four
dimensions and it corresponds to the fifth--dimensional gauge freedom
associated with $A_5^a$.

But $\pi^a$ is a pseudoGoldstone boson; the symmetry corresponding to it is
explicitly broken by the weak $SU(m)_k$ interactions. What does dimensional
deconstruction tells us about its mass? As ACG state, the higher dimensional
gauge invariance forbids contributions to the mass of $A_5$ from energy
scales greater than $1/R$, the inverse size of the fifth dimension. However,
gauge invariance does allow a mass term for $A_5$ from $|\CW|^2$ where $\CW =
P\exp{(i\int dx_5 A_5)}$ is the nontrivial Wilson loop around the fifth
dimension. Since $|\CW|^2$ is a nonlocal operator, it cannot be generated
with a UV--divergent coefficient. On the discretized circle, $\CW = \Tr
[\Pi_{k=1}^N \exp{(ia A_{5k})}]$. In the 4--dimensional theory this is just
the gauge--invariant ${\Tr}(U_1 U_2\cdots U_N)$, and so this is what provides
the mass for $\pi^a$.  Standard power counting indicates that the strength of
$|{\Tr}(U_1 U_2\cdots U_N)|^2$ is $\Lambda^2 f^2 (g^2/16\pi^2)^N$. This is
correct only for $N=1$. For $N \ge 2$ infrared singularities from the gauge
boson masses at $g \ra 0$ overrule this power counting. ACG show this using
the Coleman--Weinberg potential for $\pi^a$. Contributions to the mass for
$N=2$ come from the infrared to the ultraviolet regions, so that $M^2 \propto
g^4 f^2 \log(\Lambda^2/\CM^2_B) \sim g^4 f^2 \log(4\pi^2 N^2/g^2)$ where
$\CM^2_B \sim g^2 f^2 /N^2$ is a typical $SU(m)$ gauge boson mass; for $N \ge
3$ the IR region dominates and $M^2 \propto g^4 f^2$.

The same dependence of the $g^2$--power on $N$ is readily seen by calculating
$M^2$ from Dashen's formula~\cite{rfd}:
\be\label{eq:pcaca}
4f^2 M^2 \delta_{ab} = i^2\langle\Omega|[Q^a_\pi,[Q^b_\pi,\CH_N]]
|\Omega\rangle
\ee
The $\pi^a$ chiral symmetry breaking Hamiltonian $\CH_N$ is depicted in
Fig.~3 and is given by
\be\label{eq:HN}
\CH_N \simeq i^{N+1} g^{2N} \int{d^4q\over{(2\pi)^4}}
\biggl({1\over{q^2}}\biggr)^N \ts \int \sum_{\{c_l\} = 1}^{m^2-1}
\prod_{k=1}^N \biggl[d^4 x_k \ts e^{iq\cdot x_k}  \ts g^{\mu_k\nu_k} \ts
T\biggr(j^{c_k}_{Lk\ts \mu_k}(x_k) j^{c_{k+1}}_{Rk\ts \nu_{k+1}}(0)\biggr)
\biggl] \ts,
\ee
where $j^{c}_{\lambda k\ts \mu} = \ol \psi_{\lambda k}\gamma_\mu t_c
\psi_{\lambda k}$. The infrared divergence in this Hamiltonian may be cut off
by replacing the massless gauge propagators by $\Pi_{k=1}^N \ts (q^2 -
\CM_k^2)^{-1}$. This ``round--the--world'' Hamiltonian corresponds to the
effective Wilson--loop interaction
\be\label{eq:prodU}
\CH_W = {C_W g^4\over{16\pi^2}} \sum_{\{c_l\} = 1}^{m^2-1} \prod_{k=1}^N
\Tr\left(t_{c_k}U_k \ts t_{c_{k+1}} U^\dag_k\right) =
{C_W g^4\over{2^N 16\pi^2}} \left\vert\Tr \left(U_1 U_2 \cdots
U_N\right)\right\vert^2\ts,
\ee
where $C_W = \CO(\log(4\pi^2/g^2))$ for $N=2$ and $\CO(1)$ for $N\ge 3$.


If $g^2/4\pi \sim 10^{-2}$ in this moose ring model, the PGB is much lighter
than its underlying compositeness scale $\Lambda$. Unfortunately, it cannot
be used as a light composite Higgs with $v \sim M$ because its quartic
self--interactions are all too weak, either derivatively coupled and of order
$p^4/f^4 \sim g^8$ for typical momentum $p \sim M$, or induced directly by
the weak gauge interactions as in $\CH_W$ and of order $g^4$. This is in
accord with what would be expected from dimensional deconstruction with its
$A_5$ interpretation of $\pi$. To overcome this, ACG went to a 6--dimensional
model with nonderivative PGB interactions. We consider this model in the rest
of this paper.

In Section~2 we describe ACG's model in which the condensed moose diagram is
the discretization of a torus with $SU(m)$ gauge groups (weak coupling $g$)
at $N\times N$ sites connected by nonlinear sigma model links. This
corresponds to a 6--dimensional gauge model with the fifth and sixth
dimensions compactified on the torus. In the 4--dimensional view of the
model, the Higgs mechanism gives mass to all but one of the $SU(m)$ gauge
multiplets, and several PGBs remain to get mass and mutually interact. Two of
these correspond to the fifth and sixth components of the gauge field; the
others do not have such simple topological interpretations. We discuss the
expectations from strict dimensional deconstruction for the PGB masses and
interactions and show, in particular, that the quartic interaction of the
PGBs corresponding to $A_{5,6}$ should be $\CO(g^2/N^2)$.  This is not
strong, but it may be large enough compared to $M^2/\Lambda^2$ to give $v
\sim M \ll \Lambda$. We present a UV completion of this model consisting of a
QCD--like dynamics of fermions with strong gauge interactions, just as ACG
did for the 5--dimensional moose ring model. Then, for simplicity and for its
phenomenological relevance~\cite{ac}, we restrict this model to $N=2$. It has
five composite PGB multiplets. In Section~3 we estimate the PGB masses,
identifying the structure of the leading $g^4 \log(1/g^2)$ and $g^4$
contributions to $M^2$. The $g^4 \log(1/g^2)$ terms are the same as in the
moose ring model and their form is predicted by dimensional
deconstruction. At that order, one of the PGBs {\it not} corresponding to
$A_{5,6}$ remains massless. When $\CO(g^4)$ terms are added, all five PGBs
have comparable mass. In Section~4 we consider the nonderivative interactions
of the PGBs. The interactions produced by the QCD--like UV completion have
neither the form nor the strength of those predicted by dimensional
deconstruction. In particular, the interactions are $\CO(g^4)$, too weak to
give a Higgs vev comparable to its mass. In Section~5 we study a UV
completion that adds elementary scalars interacting strongly with themselves
and the fermions. These induce the PGB strong self--interactions expected
from dimensional deconstruction. However, for $N=2$ these scalar interactions
also give large masses to {\it all} the PGBs. At the least, this changes the
low--energy phenomenology of the model; at worst, it eliminates the
candidates for a light composite Higgs. This difficulty of constructing light
composite Higgs bosons seems to be general: The desired quartic interactions
explicitly break the symmetries keeping the PGBs light. If the interactions
are strong, the PGBs are not PGBs at all, and conversely. In any case, what
happens depends critically on the condensed--moose theory's UV completion.

\begin{figure}[tb]
\vbox to 8.5cm{
\vfill
\includegraphics{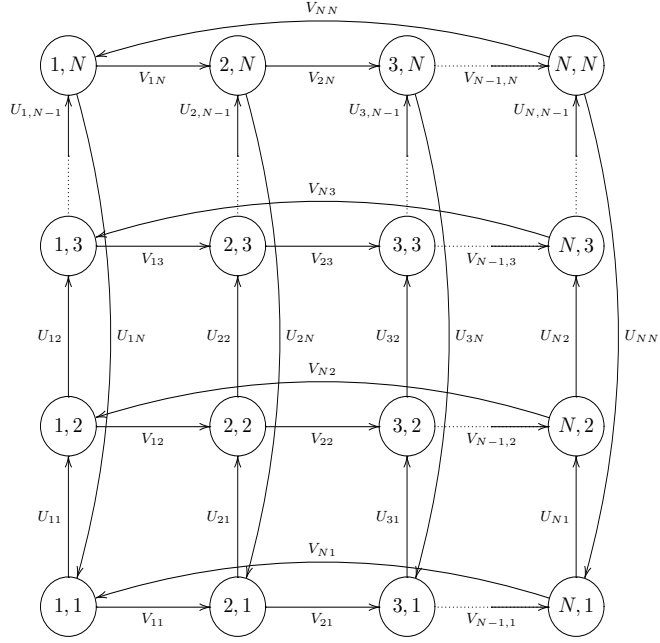}
\vfill
}
\caption{The condensed moose for the toroidal model of Ref.~\cite{acgb}. The
  weak $SU(m)_{kl}$ group is denoted by a circle at the site
  $(k,l)$. The site $(k,l)$ is identified with the sites $(k+N,l)$ and
  ($k,l+N)$. Nonlinear sigma model link--fields $U_{kl}$ and $V_{kl}$ transform
  according to Eq.~(\ref{eq:UVtransform}).
\label{fig:a}}
\end{figure}


\section*{2. The $d=6$ Toroidal Moose Model}

In Ref.~\cite{acgb}, ACG considered a model in which the condensed moose is
the discretization of a torus with $N\times N$ sites; see Fig.~4. The sites
are labeled by integers $(k,l)$ with $k$ identified with $k+N$ and $l$ with
$l+N$. The weakly--coupled gauge groups $SU(m)_{kl}$ (all with coupling $g$)
are located at the sites. The sites are linked by $U_{kl}$ and $V_{kl}$. They
connect the sites $(k,l)$ to $(k,l+1)$ and to $(k+1,l)$, respectively,
according to the $SU(m)_{kl}$ transformations
\be\label{eq:UVtransform}
U_{kl} \ra W_{kl} \ts U_{kl} \ts W^\dag_{k,l+1}\ts, \qquad 
V_{kl} \ra W_{kl} \ts V_{kl} \ts W^\dag_{k+1,l} \ts.
\ee
In the 4--dimensional theory, the link variables are nonlinear sigma model
fields involving $2N^2$ $SU(m)$ adjoints of composite Goldstone bosons,
$\pi_{u,kl} = \sum_a \pi^a_{u,kl} t_a$ and $\pi_{v,kl} = \sum_a
\pi^a_{v,kl} t_a$:
\be\label{eq:nlsigma}
U_{kl} = \exp{(i\pi_{u,kl}/f)}\ts, \qquad
V_{kl} = \exp{(i\pi_{v,kl}/f)} \ts.
\ee

The $SU(m)_{kl}$ gauge bosons eat $N^2-1$ sets of GBs. From the covariant
derivatives,
\bea\label{eq:covder}
&&D^\mu U_{kl} = \partial^\mu U_{kl} -i A^\mu_{kl} U_{kl} 
    + i U_{kl} A^\mu_{k,l+1} \ts, \nn\\
&&D^\mu V_{kl} = \partial^\mu V_{kl} -i A^\mu_{kl} V_{kl} 
    + i V_{kl} A^\mu_{k+1,l} \ts,
\eea
it is easy to determine that the mass eigenstate vector bosons and their
masses are
\bea\label{eq:Bvectors}
B^\mu_{mn} &=& \sum_{k,l} \left(\zeta^*_{(mn)}\right)_{(kl)}\ts A^\mu_{kl}
\equiv {1\over{N}}\sum_{k,l} e^{-2i(km+ln)\pi/N} \ts A^\mu_{kl} \ts,\nn\\
\CM^2_{B,mn} &=& 4g^2 f^2 \left(\sin^2 \left({m\pi\over{N}}\right)
+ \sin^2 \left({n\pi\over{N}}\right)\right) \qquad (m,n=1,\dots,N)\ts.
\eea
The massless gauge boson is $B^\mu_{NN} = N^{-1} \sum_{k,l} A^\mu_{kl}$ and
its coupling is $g/N$.

Among the $N^2+1$ leftover PGBs, two that are especially interesting are
\be\label{eq:piupiv}
\pi_u = {1\over{N}}\sum_{k,l} \pi_{u,kl} \ts, \qquad
\pi_v = {1\over{N}}\sum_{k,l} \pi_{v,kl} \ts.
\ee
These are the analogs of $\pi$ in the moose ring model, the zero modes
associated with going around the torus in the $U$ and $V$--directions. ACG
used these two PGBs as light composite Higgs multiplets.\footnote{In
Ref.~\cite{acgb}, the weak groups are all $SU(3)$ except at the (1,1) site
where the $SU(2)\otimes U(1)$ subgroup of $SU(3)$. This stratagem gets the
putative Higgses out of the adjoint and into $SU(2)$ doublets where they
belong. We shall not need to complicate our exposition by inserting a weak
gauge defect at one site.}

What do we expect for the masses and couplings of $\pi_u$ and $\pi_v$ from
dimensional deconstruction? Viewing the condensed moose as the compactified
fifth and sixth dimensions of a 6--dimensional gauge theory, the toroidal
circumference once again is $R = Na$ with $a=1/gf$, the gauge coupling is
$g_6 = ga$, and the extra--dimensional gauge fields are $A^a_{5,6} =
g\pi^a_{u,v}/N$. As in the moose ring model, dimensional deconstruction tells
us that the leading contributions to their masses come from the Wilson loops
around the fifth and sixth dimensions, e.g.,
\bea\label{eq:wilson}
|\CW_5|^2 &=& \left|P\exp{\left(i\int dx_5 A_5\right)}\right|^2 =
\left|\Tr\left\{\prod_{l=1}^N \exp{\left(ia A_{5 \ts
          kl}\right)}\right\}\right|^2 \\ 
&=&
\left|\Tr\left\{\left[\exp{\left(i\pi_u/Nf\right)}\right]^N\right\}\right|^2
= \left|\Tr\left(U_{k1} \cdots U_{kN}\right)\right|^2 \qquad(k = 1,\dots,N)
 \ts.\nn
\eea
Thus, as in the 5--dimensional model, we expect $M^2_{\pi_{u,v}} \propto g^4
f^2 \log(4\pi^2 N^2/g^2)$ for $N=2$ and $g^4 f^2$ for $N \ge 3$. The last
equality in Eq.~(\ref{eq:wilson}) assumes that $\pi_{u,v}$ are the only light
PGBs, so that $U_{kl} \cong \exp(i\pi_u/Nf)$ and $V_{kl} \cong
\exp(i\pi_v/Nf)$ at low energies. We shall see in the next section that this
is not always true; the PGB masses depend on the nature of the theory's UV
completion.

The quartic self--interactions of the PGBs of the moose ring model are weak,
at most $\CO(g^4)$. In the 6--dimensional gauge model, dimensional
deconstruction implies the existence of a stronger nonderivative interaction
corresponding to
\be\label{eq:sixdim}
\Tr F_{56}^2 = \Tr\left(\left[A_5,A_6\right]^2\right) + \cdots =
\lambda  \Tr\left(\left[\pi_u,\pi_v\right]^2\right) + \cdots
\ts.
\ee
This interaction comes from the plaquette operators~\cite{acgb}
\be\label{eq:plaq}
\CH_{\Box} = \sum_{k,l} \lambda_{kl} \ts f^4 \ts \Tr\left(U_{kl} V_{k,l+1}
U_{k+1,l}^\dag V_{kl}^\dag\right) + {\rm h.c.}
\ee
Note that $\CH_{\Box}$ does not contribute to the $\pi_{u,v}$ masses. The
strength of the $\Tr([\pi_u,\pi_v]^2)$ term is $\lambda = \half\sum_{k,l}
\lambda_{kl}/N^4$. The $\lambda_{kl}$ are fixed by dimensional deconstruction
as follows:\footnote{I thank Bill Bardeen for this argument.} The
6--dimensional action including the nonderivative term in
Eq.~(\ref{eq:sixdim}) is
\be\label{eq:sixaction}
\int d^4x \ts a^2 {1\over{g_6^2}} \left[\sum_{\mu,\nu=1}^4 F_{\mu\nu}^2 
    + {1\over{a^4}} \sum_{k,l} \Tr\left(U_{kl} V_{k,l+1}
      U_{k+1,l}^\dag  V_{kl}^\dag\right) + {\rm mixed \,\, terms} \right]
\ts.
\ee
This gives $\lambda_{kl} = (f^4g_6^2a^2)^{-1} = g^2$ and
\be\label{eq:DDquartic}
\lambda = {g^2\over{2N^2}} \ts.
\ee
This may or may not be large enough to give a Higgs vev $v$ comparable to
$M_{\pi_{u,v}}$, depending on the $N$--dependence of the Higgs masses. 

The question of whether we can realize Eq.~(\ref{eq:DDquartic}) is a major
focus of this paper. We pose the question as follows: Can we construct a UV
completion of the condensed moose that generates $\Tr([\pi_u,\pi_v]^2)$ with
strength $\CO(g^2/N^2)$. Below we present a QCD--like UV completion.  We find
that this produces $\lambda =\CO(g^4)$, and a vev of $\CO(\Lambda)$. In
Section~5 we study a completion involving strongly interacting elementary
scalars. It also fails to produce the desired hierarchy of $M$, $v$, and
$\Lambda$. While we have not proved that no UV completion exists which
realizes the expectation of deconstruction, we expect this is so. In any
event, the outcome of the low--energy Higgs theory depends crucially on its
UV completion.

The simplest UV completion of this model, and the one we shall adopt in the
next two sections, follows the strategy of the moose ring model and is based
on QCD--like dynamics at the scale $\Lambda \simeq 4\pi f$. Since the link
variables involve $2N^2$ Goldstone bosons $\pi_{u,kl}$ and $\pi_{v,kl}$, we
assume there are $2N^2$ strongly coupled $SU(n)$ gauge groups. These are
located at sites $(k,l+\half)$ and $(k+\half,l)$ for $k,l = 1,\dots,N$. The
strongly interacting massless fermions of this model are
\bea\label{eq:fermionsb}
&& \psi_{R\ts (k,l+\half)} \in (n,m,1)\ts, \ts\ts\ts \psi_{L\ts (k,l+\half)}
\in (n,1,m) \ts\ts\ts {\rm of} \ts\ts\ts
(SU(n)_{k,l+\half}\ts,SU(m)_{kl}\ts,SU(m)_{k,l+1}) \ts;\nn\\ 
&& \\
&& \psi_{R\ts (k+\half,l)} \in (n,m,1)\ts, \ts\ts\ts \psi_{L\ts (k+\half,l)}
\in (n,1,m) \ts\ts\ts {\rm of} \ts\ts\ts
(SU(n)_{k+\half,l}\ts,SU(m)_{kl}\ts,SU(m)_{k+1,l}) \ts.\nn
\eea
The UV--completed toroidal moose is shown in Fig.~5.

\begin{figure}[tb]
\vbox to 8.5cm{
\vfill
\includegraphics{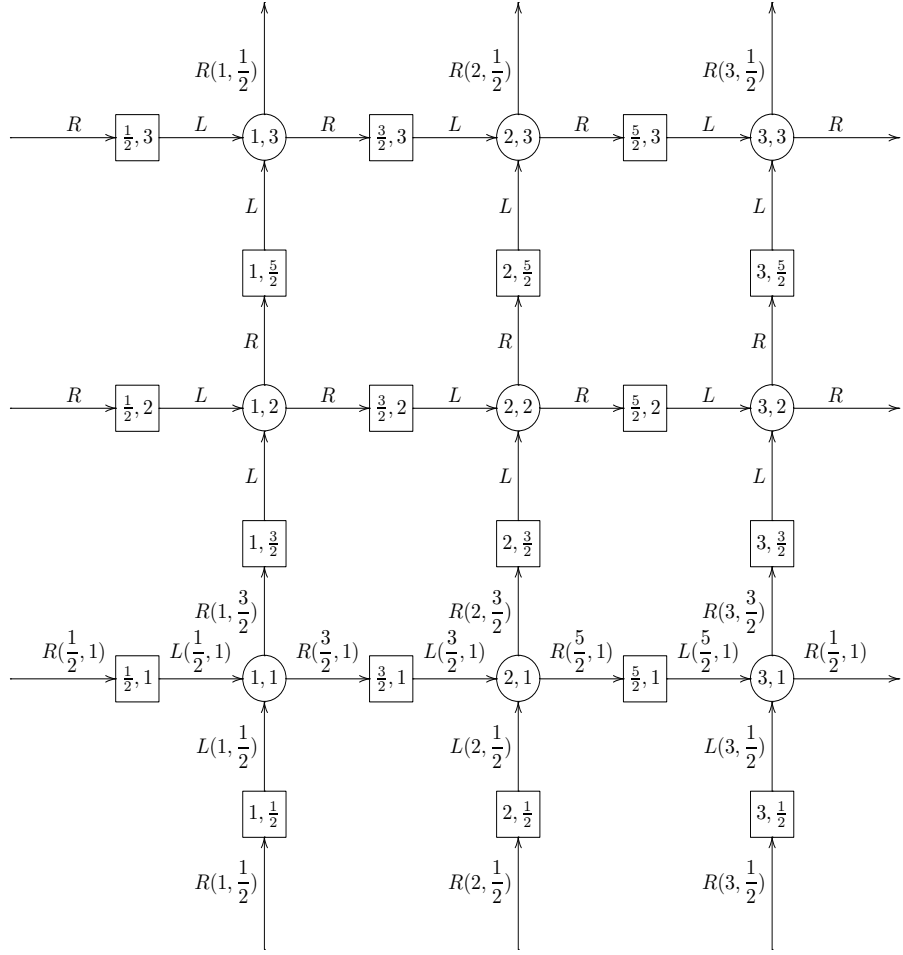}
\vfill
}
\vskip 1.75in
\caption{The complete moose for the toroidal model of Ref~\cite{acgb} with a
  QCD--like UV completion. The weak $SU(m)_{kl}$ gauge groups are as in
  Fig.~4, and the strong $SU(k-\half,l)$ and $SU(k,l-\half)$
  gauge groups are indicated by squares. Fermions transform as in
  Eq.~(\ref{eq:fermionsb}).
}
\end{figure}


\begin{figure}[tb]
\vbox to 7cm{
\vfill
\includegraphics{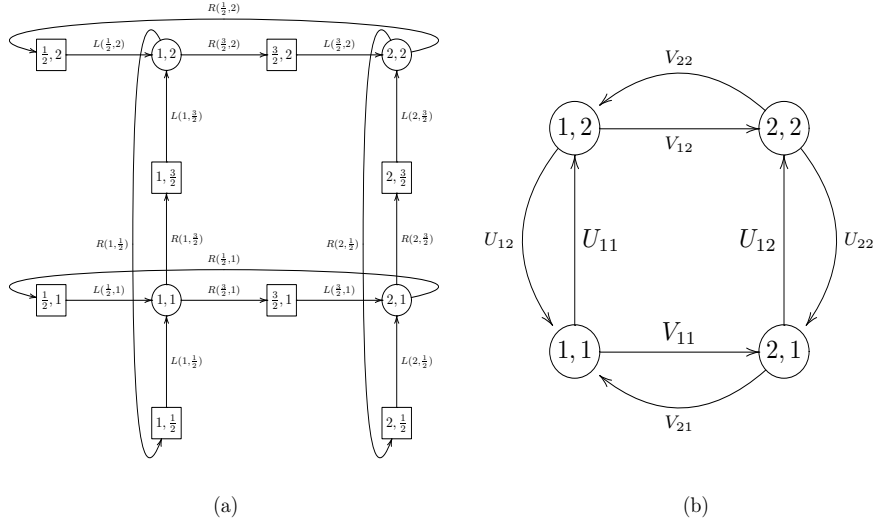}
\vfill
}
\caption{The full (a) and condensed (b) mooses for the $N=2$ toroidal model
  of Ref.~\cite{acgb} with a QCD--like UV completion. Notation is as in
  Figs.~4 and~5.
\label{fig:a}}
\end{figure}


These fermions interactions are invariant under an $[SU(m)_L \otimes
SU(m)_R]^{2N^2}$ chiral symmetry. Strong $SU(n)$ dynamics cause the
condensates 
\bea\label{eq:condensates}
&&\langle\Omega | \ol \psi_{L\ts (k,l+\half)} \psi_{R\ts (m,n+\half)}
|\Omega\rangle = -\delta_{km}\ts \delta_{ln} \Delta  \longleftrightarrow
4\pi f^3 U_{kl}\ts \delta_{km}\ts \delta_{ln} \ts,\nn\\
&&\langle\Omega | \ol \psi_{L\ts (k+\half,l)} \psi_{R\ts (m+\half,n)} 
|\Omega\rangle = -\delta_{km}\ts \delta_{ln} \Delta  \longleftrightarrow 4\pi
f^3 V_{kl}\ts \delta_{km}\ts \delta_{ln} \ts, 
\eea
so that this symmetry breaks spontaneously to the diagonal $[SU(m)_V]^{2N^2}$
subgroup with the appearance of the $\pi_{u,v\ts kl}$.

From now on, we restrict ourselves to the case $N=2$, partly for the
phenomenological reason noted earlier and partly because of its simplicities
and peculiarities. The full and condensed $N=2$ mooses are shown in
Fig.~6. Note that every pair of adjacent lattice sites in the condensed moose
are connected by two link variables, $U_{kl}$ and $U_{k,l+1}$ or $V_{kl}$ and
$V_{k+1,l}$. The gauge boson masses are $\CM^2_{11} = 8g^2f^2$, $\CM^2_{12} =
\CM^2_{21} = 4g^2f^2$, and $\CM^2_{22} = 0$.

From Eq.~(\ref{eq:Bvectors}) and the fermion $SU(m)_{kl}$ current,
\be\label{eq:Aklcurrent}
j^a_{\mu, kl} = j^a_{R\mu \ts(k,l+\half)} + j^a_{L\mu
\ts(k,l-\half)} + j^a_{R\mu \ts(k+\half,l)} + j^a_{L\mu \ts(k-\half,l)} \ts,
\ee
we read off the Goldstone boson eaten by $B^\mu_{mn}$:
\be\label{eq:eatengbs}
\pi^a_{(mn)} = \half\sum_{k,l=1}^2
\left(\zeta_{(mn)}\right)_{(kl)} [\pi^a_{u,kl} - \pi^a_{u,\ts k,l-1} +
\pi^a_{v,kl} - \pi^a_{v,\ts k-1,l}] \ts.
\ee
A convenient basis for the five physical GBs, whose masses and couplings we
will estimate in the next two sections, is:
\bea\label{eq:pgbs}
\pi_u &=& \half\left[\pi_{u,11} + \pi_{u,12} + \pi_{u,21} +
  \pi_{u,22}\right] \nn\\ 
\pi_v &=& \half\left[\pi_{v,11} + \pi_{v,21} + \pi_{v,12} +
  \pi_{v,22}\right] \nn\\
\pi'_u &=& \half\left[\pi_{u,11} + \pi_{u,12} - \pi_{u,21} -
  \pi_{u,22}\right] \nn\\
\pi'_v &=& \half\left[\pi_{v,11} + \pi_{v,21} - \pi_{v,12} -
  \pi_{v,22}\right] \nn\\
\pi'_{uv} &=& \half\left[\pi_{u,11} + \pi_{u,22} - \pi_{u,12} -
  \pi_{u,21} -
\left(\pi_{v,11} + \pi_{v,22} - \pi_{v,12} -
  \pi_{v,21}\right)\right] \ts.
\eea
%
\nin The inverse transformations, which are useful when expanding plaquette
interactions, are:
\bea\label{eq:inverse}
\pi_{u,11} &=& \half\left[\pi_u + \pi'_u + \pi_{(21)} +
\textstyle{1\over{\sqrt{2}}}\left(\pi'_{uv} + \pi_{(11)}\right)\right] \nn\\
\pi_{u,12} &=& \half\left[\pi_u + \pi'_u - \pi_{(21)} -
\textstyle{1\over{\sqrt{2}}}\left(\pi'_{uv} + \pi_{(11)}\right)\right] \nn\\
\pi_{u,21} &=& \half\left[\pi_u - \pi'_u + \pi_{(21)} -
\textstyle{1\over{\sqrt{2}}}\left(\pi'_{uv} + \pi_{(11)}\right)\right] \nn\\
\pi_{u,22} &=& \half\left[\pi_u - \pi'_u - \pi_{(21)} +
\textstyle{1\over{\sqrt{2}}}\left(\pi'_{uv} + \pi_{(11)}\right)\right]\ts;
\nn\\ \\
\pi_{v,11} &=& \half\left[\pi_v + \pi'_v + \pi_{(12)} -
\textstyle{1\over{\sqrt{2}}}\left(\pi'_{uv} - \pi_{(11)}\right)\right] \nn\\
\pi_{v,21} &=& \half\left[\pi_v + \pi'_v - \pi_{(12)} +
\textstyle{1\over{\sqrt{2}}}\left(\pi'_{uv} - \pi_{(11)}\right)\right] \nn\\
\pi_{v,12} &=& \half\left[\pi_v - \pi'_v + \pi_{(12)} +
\textstyle{1\over{\sqrt{2}}}\left(\pi'_{uv} - \pi_{(11)}\right)\right] \nn\\
\pi_{v,22} &=& \half\left[\pi_v - \pi'_v - \pi_{(12)} -
\textstyle{1\over{\sqrt{2}}}\left(\pi'_{uv} - \pi_{(11)}\right)\right]
\ts. \nn
\eea

\section*{3. PseudoGoldstone Boson Masses}

As in the moose ring model for $N=2$, the leading $g^4\log(1/g^2)$
contribution to the PGB masses comes from four distinct round--the--world
graphs of the type shown in Fig.~7. The Hamiltonian, analogous to $\CH_N$ in
Eq.~(\ref{eq:HN}), is
\bea\label{eq:HTtwo}
\CH_2 &\simeq& i g^4 \int{d^4q\over{(2\pi)^4}}
\biggl({1\over{q^2}}\biggr)^4\ts \int d^4x d^4y \ts e^{iq\cdot(x+y)}
\ts g^{\mu\nu} g^{\lambda\rho} \\
&\times& \sum_{c,d = 1}^{m^2-1} \sum_{k=1}^2
\left\{T\Bigl(j^{c}_{R\ts \mu (k,\half)}(x)
               j^{d}_{L\ts \lambda(k,\half)}(0)\Bigr) \ts 
T\Bigl(j^{d}_{R\ts \rho (k,\threehalves)}(y)
        j^{c}_{L\ts \nu(k,\threehalves)}(0)\Bigr) + (k,l/2)
\leftrightarrow(l/2,k)\right\}
\ts.\nn
\eea 
This corresponds to the effective interaction
\bea\label{eq:Htwo}
\CH_2 &=& -{C_2 f^4 g^4\over{32\pi^2}} \ts \log\left({4\pi^2\over{g^2}}\right)
\sum_{c,d = 1}^{m^2-1}\sum_{k,l=1}^2 
\biggl\{\Tr\bigl(t_c U_{kl} \ts t_d U^\dag_{kl}\bigr)\ts
\Tr\bigl(t_c U^\dag_{k,l+1} \ts t_d U_{k,l+1}\bigr) +
\bigl(U_{kl} \ra V_{lk} \bigr) \biggr\}\nn\\
&=& -{C_2 f^4 g^4\over{128\pi^2}} \ts \log\left({4\pi^2\over{g^2}}\right)
\sum_{k,l=1}^2 \biggl\{\Bigl|\Tr\bigl(U_{kl} U_{k,l+1}\bigr)\Bigr|^2 +
(U_{kl} \ra V_{lk})\biggr\}
\eea
where we will see that $C_2 \simeq 6$.

We can estimate the IR--singular contributions to the PGB masses by the
ancient method of current algebra combined with Weinberg's spectral function
sum rules~\cite{pcac,sfsr}. As in QCD, we assume the vector--axial vector
spectral function $\Delta_{VA}$ can be saturated with with a massless
pseudoscalar and a single vector and axial vector meson of masses $M_{V,A}
\simeq \Lambda$ and dimensionless couplings $f_{V,A}$ to the $(V,A)$ currents
gives
\bea\label{eq:delVAb}
\Delta_{VA} &=& {f_V^2 M_V^2\over{q^2-M_V^2}} - {f_A^2 M_A^2\over{q^2-M_A^2}} -
{4f^2\over{q^2}} \nn\\
&=& {f_V^2 M_V^4\over{q^2}} \left({1\over{q^2-M_V^2}} -
  {1\over{q^2-M_A^2}}\right) \ts.
\eea
The second equality follows from the spectral function sum rules,
\bea\label{eq:sfsreq}
&& f_V^2 M_V^2 - f_A^2 M_A^2 = 4f^2 \ts;\nn\\
&& f_V^2 M_V^4 - f_A^2 M_A^4 = 0 \ts.
\eea
We obtain 
\be\label{eq:IRmasses}
\left(M^2_{\pi_{u,v}}\right)_{g^4\log g^2} = 
\left(M^2_{\pi'_{u,v}}\right)_{g^4\log g^2} \simeq
  {3m g^4 f^2\over{32\pi^2}} \log\left({4\pi^2\over{g^2}}\right) \ts,
\ee
where we put $\log(M^2_V/\CM^2_B) = \log(4\pi^2/g^2)$ for a typical $SU(m)$
gauge mass of $4g^2f^2$ and used $f^2_V M^2_V(M^2_A - M^2_V)/4f^2 = 1$ from
Eq.~(\ref{eq:sfsreq}). We read off
\be\label{eq:ctwo}
C_2 \simeq 6 \ts.
\ee
All mixing terms vanish and $(M^2_{\pi'_{uv}})_{g^4\log g^2} = 0$. In other
words, dimensional deconstruction again predicts the origin but not the
magnitude of the leading contribution to the $\pi_{u,v}$ masses. It fails to
mention the $\pi'_{u,v}$ and the fact that they are degenerate with
$\pi_{u,v}$. It completely misses the $\pi'_{uv}$ and its masslessness at
order $g^4\log(1/g^2)$.

\begin{figure}[tb]
\vbox to 4cm{
\vfill
\includegraphics{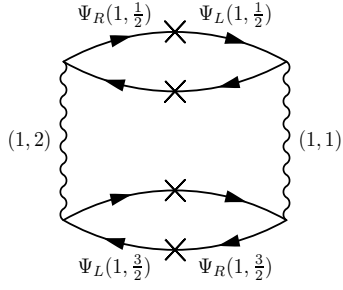}
\vfill
}
\caption{Graphical depiction of a typical term in the Hamiltonian $\CH_2$ in
Eq.~(\ref{eq:HTtwo}). Fermions transform as indicated in Fig.~6a under the
weak gauge groups $SU(m)_{kl}$, whose bosons are identified in the figure. An
$\times$ indicates a dynamical mass insertion. As in Fig.~3, strong $SU(n)$
gauge boson interactions within each fermion loop are not indicated and there
are no strong gauge interactions between loops.
\label{fig:a}}
\end{figure}

\begin{figure}[tb]
\vbox to 9cm{
\vfill
\includegraphics{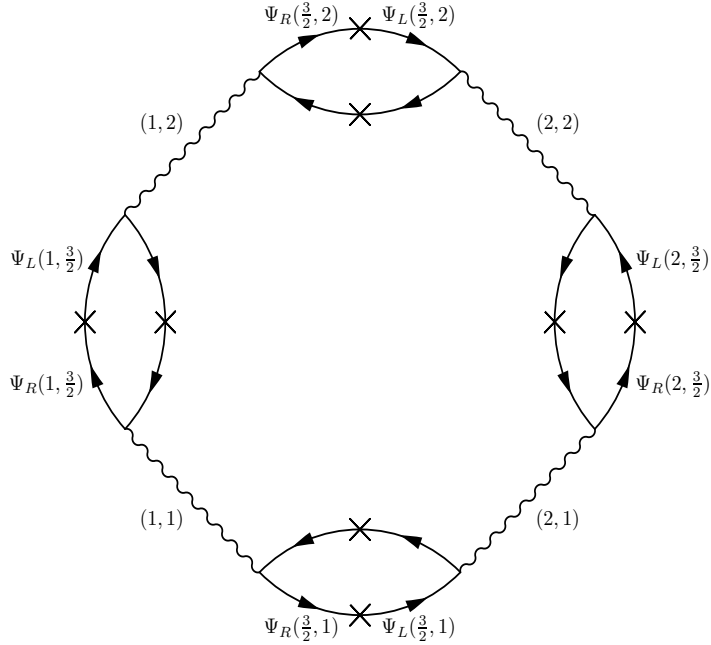}
\vfill
}
\caption{Typical graph contributing to the Hamiltonian $\CH_4$ in
  Eq.(\ref{eq:Hfour}). The notation is as in Fig.~7.
\label{fig:a}}
\end{figure}


All the PGBs, including $\pi'_{u,v}$, get masses from 16 one--loop graphs of
the type shown in Fig.~8. Because of infrared singularities, these are
actually $\CO(g^4)$. They are represented by the $SU(m)$--invariant effective
Hamiltonian
\bea\label{eq:Hfour}
\CH_4 &=& -{C_4 \ts g^4f^4\over{16\pi^2}} \sum_{k,l=1}^2 \ts \biggl\{
\Bigl|\Tr\bigl(U_{kl} V_{k,l+1} U^\dag_{k+1,l} V^\dag_{kl}\bigr)\Bigr|^2
\nn\\ 
&&\qquad\qquad\qquad +\ts  \Bigl|\Tr\bigl(U_{kl} V_{k,l+1} U_{k+1,l+1}
V^\dag_{kl}\bigr)\Bigr|^2
+\ts \Bigl|\Tr\bigl(U_{kl} V_{k,l+1} U^\dag_{k+1,l} V_{k+1,l}\bigr)\Bigr|^2 \\
&&\qquad\qquad\quad\ts + \ts \half \ts \Bigl|\Tr\bigl(U_{kl} V_{k,l+1}
U_{k+1,l+1} V_{k+1,l}\bigr)\Bigr|^2 
+ \ts \half \ts \Bigl|\Tr\bigl(U_{kl} V^\dag_{k+1,l+1} U_{k+1,l+1}
V^\dag_{kl}\bigr)\Bigr|^2 \biggr\} \ts.\nn
\eea
Note that these terms are invariant under the interchanges $U_{kl}
\leftrightarrow V_{lk}$. The first four terms in $\CH_4$ are the sum of the
{\it squares} of the plaquette interaction $\CH_{\Box}$ in
Eq.~(\ref{eq:plaq}). The other terms are allowed by gauge invariance for
$N=2$ and have exactly the same strength.\footnote{This $N=2$ case is
special. In an $N\times N$ toroidal lattice with periodic boundaries, there
are $N^2$ plaquettes for $N\ge 3$.} We will discuss in Section~4 why the
linear plaquette interaction does not appear.

The PGB masses from $\CH_4$ are easily evaluated. There is no mixing in the
sum of the terms and we find:
\bea\label{eq:PGBmasses}
&& \left(M^2_{\pi_u}\right)_{g^4} = \left(M^2_{\pi_v}\right)_{g^4} = {m C_4
\ts  g^4 f^2\over{2\pi^2}} \ts,
\nn\\
&& \left(M^2_{\pi'_u}\right)_{g^4} = \left(M^2_{\pi'_v}\right)_{g^4} = {3m
C_4 \ts g^4 f^2\over{4\pi^2}} \ts,
\nn\\
&& \left(M^2_{\pi'_{uv}}\right)_{g^4} = {2m C_4 \ts g^4 f^2\over{\pi^2}}
\ts.
\eea
We estimate $C_4$ by replacing the four massless gauge propagators in Fig.~8
by the mass eigenstate product $\Pi_{k,l}\ts (q^2-\CM^2_{kl})^{-1}$ and using
the spectral functions $\Delta_{VA}$:
\be\label{eq:cfour}
C_4 \simeq {3\over{16}} \left(1-\log{2}\right) \ts.
\ee
It is easy to see from Eq.~(\ref{eq:Hfour}) where the ratio
$(M^2_{\pi_{u,v}})_{g^4} : (M^2_{\pi'_{u,v}})_{g^4} :
(M^2_{\pi'_{uv}})_{g^4} = 8:12:32$ comes from. Just put all PGB fields
but the one in question to zero and expand to $\CO((\pi)^2)$.

Let us estimate the ratio of the two contributions to $M^2_{\pi_u}$. We take
$g^2/4\pi = 1/30$. Then $(M^2_{\pi_u})_{g^4\log g^2}/(M^2_{\pi_u})_{g^4}
\simeq 15 \simeq 2(M^2_{\pi_{uv}})_{g^4}/(M^2_{\pi_u})_{g^4}$; i.e., all the
PGBs have roughly the same mass. With this QCD--like UV completion of the
toroidal moose model, then, there are {\it five} sets of light PGBs and a
very rich phenomenology. This does not change if there is a gauge defect with
$SU(2) \otimes U(1)$ at site $(1,1)$. Nor does the situation change
qualitatively for $N>2$. In sum, the particle spectrum of the 6--dimensional
gauge theory is not a very good representation of the 4--dimensional one at
energies well below $\Lambda.$

\section*{4. PseudoGoldstone Self--Interactions}

Deconstructing the 6--dimensional toroidal moose led us to expect the
nonderivative interaction $\Tr([\pi_u,\pi_v])^2$ with strength $g^2/N^2$. It
was to come from the lattice version of $\int g^{-2}_6 \ts \Tr F^2_{56}$,
namely, $g^2\sum_{k,l}\Tr(U_{kl} V_{k,l+1} U_{k+1,l} V^\dag_{kl})$. Instead,
our QCD--style UV completion of the model produced the squared, not the
linear, plaquette interaction $\CH_4$. Its $\CO(g^4)$ coupling is too weak
to produce a light composite Higgs vev much less than $\Lambda$.

The linear interaction does not appear for {\it any} $N$ because this theory
is invariant under reflection of any and all fermion fields $\psi_{L,R}$ and,
hence, under the reflection of any $U_{kl}$ and any $V_{kl}$. This by itself
does not preclude the existence of $\Tr([\pi_u,\pi_v])^2$, which indeed
appears in the expansion of $|\Tr(UVU^\dag V^\dag)|^2$. However, for the
phenomenologically interesting case of $N=2$, even this interaction and many
others like it are absent, at least to $\CO(g^4)$. This follows from the
invariance of $\CH_4$ under the replacement of any single $U_{kl}$ by
$U^\dag_{k,l+1}$ or $V_{kl}$ by $V^\dag_{k+1,l}$. Thus, interactions arising
from $\CH_4$ are invariant under any {\it one} replacement of the type
\bea\label{eq:replace}
\pi_u \pm \pi'_u \pm \textstyle{1\over{\sqrt{2}}}\pi'_{uv} &\lra&
-\left(\pi_u \pm \pi'_u\right) \pm \textstyle{1\over{\sqrt{2}}}\pi'_{uv} \ts;
\nn\\
\pi_v \pm \pi'_v \pm \textstyle{1\over{\sqrt{2}}}\pi'_{uv} &\lra&
-\left(\pi_v \pm \pi'_v\right) \pm \textstyle{1\over{\sqrt{2}}}\pi'_{uv} \ts.
\eea
Eight $\Tr([\pi_u,\pi_v])^2$ terms in $\CH_4$ are canceled by eight
$\Tr([\pi_u,\pi_v][\pm \pi_u,\mp \pi_v])$ terms.

This is not to say that all quartic PGB interactions in $\CH_4$ vanish; they
don't. But for our UV completion of the $N=2$ toroidal moose, the term
expected from dimensional deconstruction just isn't there. This is an
artifact of $N=2$; $\Tr([\pi_u,\pi_v])^2$ does appear for higher $N$. But its
coupling and that of all other nonderivative quartic interactions are still
$\CO(g^4)$. In the next section, we change the UV completion of the model to
obtain stronger quartic interactions.

\section*{5. Stronger Interactions from Elementary Scalars}

Linear plaquette interactions of any strength can be obtained by adding
strongly--interacting scalar fields to the UV completion of the toroidal
moose model.\footnote{Andy Cohen and Howard Georgi separately mentioned to me
that scalars can induce large plaquette interactions. The implementation used
here was suggested to me by Sekhar Chivukula. It is similar in spirit to
Elizabeth Simmons' model in which the gauge bosons of extended technicolor
are replaced by scalars~\cite{ehsscalar}. Elementary scalars by themselves
make the model unnatural, so the original motivation of a naturally light
composite Higgs is lost. I assume this can be fixed by supersymmetry.}  We
introduce eight complex scalar field multiplets, $\phi_{u,kl}$ and
$\phi_{v,kl}$ for $k,l=1,2$, all of which are strong $SU(n)$ singlets:
\bea\label{eq:phiuv}
&&\phi_{u,kl} \equiv {1\over{\sqrt{2m}}}\ts  \phi^0_{u,kl} +
\sum_{a=1}^{m^2-1} \phi^a_{u,kl} \ts t_a \in (m,\ol m)  \ts\ts\ts {\rm of}
\ts\ts\ts SU(m)_{kl} \otimes SU(m)_{k,l+1} \ts,\nn\\ 
&&\phi_{v,kl} \equiv {1\over{\sqrt{2m}}}\ts \phi^0_{v,kl} +
\sum_{a=1}^{m^2-1} \phi^a_{v,kl} \ts t_a \in (m,\ol m)  \ts\ts\ts {\rm of}
\ts\ts\ts SU(m)_{kl} \otimes SU(m)_{k+1,l} \ts;
\eea
i.e., the $\phi_{u,kl}$ transform like $U_{kl}$ and the $\phi_{v,kl}$ like
$V_{kl}$. To maintain equality of the weak $SU(m)_{kl}$ couplings $g$, we
require all scalar interactions to be site--symmetric. We also impose
symmetry under the interchanges $\phi_{u,kl} \leftrightarrow
\phi_{v,lk}$. This preserves the $U_{kl} \leftrightarrow V_{lk}$ symmetry ACG
needed to avoid large tree--level Higgs masses in their model with an
$SU(2)\otimes U(1)$ gauge defect at site $(1,1)$~\cite{acgb}. These
symmetries simplify our discussion; e.g., $\phi_{u,kl}$ and $\phi_{v,kl}$
have equal masses and Yukawa couplings. We assume the scalars are heavy, with
mass $M_\phi \sim \Lambda$, the strong interaction scale of the fermions.

The Yukawa interactions consistent with gauge and other symmetries
are
\be\label{eq:yukawa}
\CL_Y = \sum_{k,l=1}^2 \left[\Gamma_\phi \left( \ol \psi_{L\ts (k,l+\half)}
\ts\phi^\dag_{u,kl}\ts \psi_{R\ts (k,l+\half)} +
\ol \psi_{L\ts (k+\half,l)} \ts\phi^\dag_{v,kl} \ts\psi_{R\ts (k+\half,l)}
\right) + {\rm h.c.} \right]\ts. 
\ee
We assume $\Gamma_\phi = \Gamma^*_\phi = \CO(1)$. In the neglect of the weak
$SU(m)$ gauge interactions, this theory is still invariant under $[SU(m)_L
\otimes SU(m)_R]^{2N^2}$ with the symmetry extended to include the scalars.

When the strong $SU(n)$ interactions generate fermion condensates, the Yukawa
interactions induce a vacuum expectation value for the scalars:
\be\label{eq:phivev} \sqrt{2} \ts v_\phi \equiv \langle{\rm
Re}(\phi^0_{u,kl})\rangle = \langle{\rm Re}(\phi^0_{v,kl})\rangle =
{\Gamma_\phi\over{M^2_\phi}} \ts \langle\ol \psi_{L\ts (k,l+\half)}
\psi_{R\ts (k,l+\half)} \rangle \sim f \ts.
\ee
The chiral symmetry is again spontaneously broken to the diagonal
$[SU(m)_V]^{2N^2}$, and the Goldstone bosons are
\be\label{eq:GBs}
\Pi^a_{u,v,\ts kl} = {f \ts \pi^a_{u,v,\ts kl} + v_\phi \ts {\rm
    Im}(\phi^a_{u,v,\ts kl})\over{\sqrt{f^2+v^2_\phi}}} \ts.
\ee
Now, $U_{kl} = \exp{(i\Pi_{u,kl}/F)}$ and $V_{kl} = \exp{(i\Pi_{v,kl}/F)}$,
where $F = \sqrt{f^2+v^2_\phi}$. The eaten and physical Goldstone bosons are
the same combinations as in Eqs.~(\ref{eq:eatengbs},\ref{eq:pgbs}). If the
$\phi_{u,v}$ have strong self--interactions, then so do the PGBs, both
directly through the $\phi^4$ terms and through the plaquette interactions
they induce. The price for this will be large PGB masses.

To generate $\Tr(U_{kl} V_{k,l+1} U_{k+1,l}^\dag V_{kl}^\dag)$ and the
squared commutator expected from dimensional deconstruction, we suppose there
exists the $SU(m)_{kl}$ gauge--invariant Hamiltonian
\be\label{eq:Hphione}
\CH_{\phi 1} = \rho_1 \sum_{k,l}\Tr\bigl(\phi_{u,kl} \ts\phi_{v,\ts
  k,l+1} \ts\phi^\dag_{u,\ts k+1,l} \ts\phi^\dag_{v,kl}\bigr) + {\rm h.c.}
\ee
This produces quartic PGB interactions and, because there now need be no
reflection symmetry to forbid it, it produces $\CH_{\Box}$ in
Eq.~(\ref{eq:plaq}) with equal strengths $\lambda_{kl} \sim \rho_1$.
Deconstruction does not fix the magnitude of $\rho_1$, so we can take it to
be anything we want. In particular, we can choose $\rho_1 = \CO(g^2)$ to make
deconstruction's prediction come true. If that gives a Higgs vev too much
larger than its mass, we can just as well choose $\rho_1 = \CO(1)$.

In the $N=2$ model, however, there's more. There can be interactions that
induce the other plaquettes in $\CH_4$, but linearized:
\bea\label{eq:Hphi}
\CH_{\phi 2} &=& \rho_2 \sum_{k,l}\Bigl[\Tr\bigl(\phi_{u,kl} \ts\phi_{v,\ts
  k,l+1} \ts\phi_{u,\ts k+1,l+1} \ts\phi^\dag_{v,kl}\bigr) + 
  \Tr\bigl(\phi_{u,kl} \ts\phi_{v,\ts k,l+1}
\ts\phi^\dag_{u,\ts k+1,l} \ts\phi_{v,\ts k+1,l}\bigr)\Bigr] + {\rm h.c.}\nn\\
\CH_{\phi 3} &=& \half\rho_3 \sum_{k,l}\Tr\bigl(\phi_{u,kl} \ts\phi_{v,\ts
  k,l+1} \ts\phi_{u,\ts k+1,l+1} \ts\phi_{v,\ts k+1,l}\bigr) + {\rm h.c.}\nn\\
\CH_{\phi 4} &=& \half\rho_4 \sum_{k,l}\Tr\bigl(\phi_{u,kl}
\ts\phi^\dag_{v,\ts k+1,l+1} \ts\phi_{u,\ts k+1,l+1} \ts\phi^\dag_{v,\ts
  kl}\bigr) + {\rm h.c.} \ts.
\eea
There can also be ``Wilson--loop'' interactions that induce the terms in
$\CH_2$ and more:\footnote{Quadratic Wilson--loop interactions can be
forbidden by discrete symmetries of the type discussed below.}
\bea\label{eq:philoop}
\CH_{\phi 5} &=& \rho_5 \sum_{k,l} \Bigl[\Bigl|\Tr(\phi_{u,kl}\ts\phi_{u,\ts
  k,l+1}) \Bigr|^2 + \Bigl|\Tr(\phi_{v,kl}\ts\phi_{v,\ts k+1,l}) \Bigr|^2
\Bigl]\nn\\
\CH_{\phi 6} &=& \sum_{k,l,m,n} \Bigl[\rho_6 \Tr(\phi_{u,kl}\ts\phi_{u,\ts
  k,l+1}) \ts \Tr(\phi^\dag_{v,mn}\ts\phi^\dag_{v,\ts m+1,n}) + \cdots \Bigl]
+ {\rm h.c.}
\eea
Dimensional deconstruction does not fix the strengths of these interactions
either, but absent a symmetry to prevent them, there is no reason for them to
be much smaller than $\rho_1$.

However, if all these $\phi^4$ and plaquette interactions are present, they
give large $\CO(\rho_i F^2)$ squared masses to all the PGBs. What symmetries
can we invoke to prevent them? Any new symmetry must respect the $\phi_{u,kl}
\leftrightarrow \phi_{v,lk}$ interchange. The discrete phase transformations
\be\label{eq:discrete}
\phi_{u,v\ts kl} \ra \eta_{u,v\ts kl} \ts \phi_{u,v\ts kl} \ts; \qquad 
\eta_{u,kl} = \eta_{v,lk}\ts, \quad |\eta_{u,v\ts kl}|=1 \ts,
\ee
can forbid all the $\phi^4$ interactions, including $\CH_{\phi 1}$, {\it
except} $\CH_{\phi 4}$ and $\CH_{\phi 5}$. Still, these two and the effective
interactions they induce are sufficient to generate large $M^2$ terms for all
the PGBs. This illustrates what seems to be a general rule: If the PGBs have
strong self--interactions, then there is large explicit symmetry breaking and
large PGB masses. If the masses are kept small, then the self--interactions
are weak. In either case, the Higgs vev is always large, $\CO(\Lambda)$.

Finally, we might well ask why we needed the fermions $\psi_{L,R}$ in this UV
completion. Their only useful purpose was to induce the vev $v_\phi$ for the
scalars. Presumably this could have been accomplished by a negative
$M^2_\phi$.

\section*{6. Conclusions}

We conclude that, for the $N=2$ toroidal moose model at least, dimensional
deconstruction is not a reliable guide to building a model of naturally light
composite Higgs bosons. Deconstruction says the model has two light Higgs
multiplets, $\pi_u$ and $\pi_v$, one of which can be given a negative
mass--squared and vev much less than the compositeness scale $\Lambda$ by
putting an $SU(2)\otimes U(1)$ gauge defect at one site. We studied whether a
small mass and vev can be obtained with two straightforward UV completions of
the model. For the QCD--like completion, we ended up with a model containing
five light PGB multiplets and weak self--interactions so that any vev is of
order $\Lambda$. For the model which adds strongly--interacting scalars, the
five PGBs have masses and quartic couplings of $\CO(g)$ to $\CO(1)$, but any
vev is still $\CO(\Lambda)$. We could choose $\rho_1$ large and all other
scalar couplings small, but this is arbitrary, having nothing to do with
deconstruction.  Furthermore, since this model presumably requires
supersymmetry to stabilize it, it does not seem much of an advance beyond
earlier supersymmetric or technicolor models.

Finally, what about models with $N \ge 3$? With a QCD--like UV completion,
all the PGBs have roughly the same $M^2 =\CO(g^4\Lambda^2)$ and $\CO(g^4)$
quartic couplings. Adding scalars with a strong interaction $\CH_{\phi 1}$
can induce $\CH_{\Box}$, raising the mass of all PGBs except $\pi_u$ and
$\pi_v$ and giving a largeish $\Tr([\pi_u,\pi_v])^2$ coupling. The
phase--invariant interaction $\CH_{\phi 5}$ has dimension $2N$. If we include
it, its strength is naturally $\rho_5/\Lambda^{2N-4}$ with $\rho_5 \sim
1$. It induces the squared Wilson--loop interaction $\sum_k[|\Tr(U_{k1}\cdots
U_{kN})|^2 + |\Tr(V_{1k}\cdots V_{Nk})|^2]$ with strength $\rho_5 F^4
(F/\Lambda)^{2N-4}$ and $M^2_{\pi_{u,v}} \sim \rho_5 F^2 (F/\Lambda)^{2N-4}$.
Again, any Higgs vev is $v \sim \Lambda$. If we exclude $\CH_{\phi 5}$, the
squared Wilson--loop terms are generated with strength $\CO(g^4)$ by the weak
$SU(m)$ interactions. This, finally, gives a Higgs spectrum and couplings in
accord with deconstruction. Supersymmetry at the scale $\Lambda$ is still
needed to keep everything stabilized.

\section*{Acknowledgements}

I am especially indebted to Bill Bardeen, Sekhar Chivukula and Estia Eichten
for many discussions and suggestions. I have also benefitted greatly from
conversations with Nima Arkani-Hamed, Andy Cohen, Howard Georgi and Chris
Hill. I am grateful to Fermilab and its Theory Group for a Frontier
Fellowship which supported this research. It was also supported in part by
the U.S.~Department of Energy under Grant~No.~DE--FG02--91ER40676. Finally, I
thank Kevin Lynch for his valuable assistance in preparing the figures.

\vfil\eject

\end{document}